# Exploration and Research on the New Mixed-Mechanism of Superconductors

D.H. Lin · D.J. Cui · B. Li · L.F. Liu · X.Q. Wang

**Abstract** We propose a new mixed-mechanism for superconductors, which addresses not only low- but also high-, and even possible room-temperature superconductivity. We use this mixed-mechanism to explain superconductivity in different temperature regions. This new mixed-mechanism is further explored and studied by investigating the possible interactions that may form Cooper pairs.

**Keywords** Superconductivity; mixed-mechanism; Interaction.

## 1. Introduction

The low temperature superconductivity has been well studied within the framework of the conventional BCS theory. The high temperature superconductors have attracted extensive attentions, because of their high transition temperatures ($T_c$) and potential extensive applications. The basic question is whether or not the conventional electron-photon mechanism that works well for low temperature superconductivity is still applicable to the high temperature superconductivity.

Most of the present experimental results suggest that the electron-phonon mechanism still works but is no longer the dominant mechanism. We try to answer this question in this work. Many experiments show that the high temperature superconductors, mainly the cuprates, have many different properties from those of the conventional low temperature super-conductors [1-3]. These differences are the clues for us to explore and study the dynamical mechanism. It is well-known that these differences are hard to be explained completely by the BCS theory.

D.H. Lin(*) · D.J. Cui · B. Li · L.F. Liu· X.Q. Wang

College of Physics, Chongqing University, JD Duz (USA)-CQU

Institute for Superconductivity, Chongqing 400044, P.R. China.

* e-mail: lindehua_cn@aliyun.com, or dhlin@cqu.edu.cn

X.Q. Wang

State Key Laboratory of Mechanical Transmission, Chongqing University, Chongqing 400044, P.R. China.

## 2. Mixed-Mechanism for High Temperature Superconductivity

We suppose that [4], the superconductivity is determined by three single mechanism, i.e., $M_1(T,X)$, $M_2(T,X)$, and $M_3(T,X)$, which corresponds to the dominant mechanism in different temperature regions ($T$) (the low-temperature, the high-temperature, and the possible room- temperature superconductivity), respectively. For different materials ($X$), the dominant mechanism shall be also different.

We assume that the mixed-mechanism of superconductivity, $M_{mix}(T,X)$, is determined by

$$M_{mix}(T,X) \Longleftrightarrow f_1(T,X)M_1(T,X) + f_2(T,X)M_2(T,X) + f_3(T,X)M_3(T,X) \quad (1)$$

Where $f_1(T,X)$, $f_2(T,X)$, and $f_3(T,X)$ are the corresponding weight functions that vary with temperature $T$ and material $X$. These three mechanisms have their different effects (dominant or secondary) within the whole temperature region. In the low-temperature region, $M_1(T,X)$ is the dominant mechanism, while $M_2(T,X)$ and $M_3(T,X)$ play secondary role; In the high-temperature region, $M_2(T,X)$ is dominating, while $M_1(T,X)$ and $M_3(T,X)$ play the secondary roles and so on.

So far, only the mechanism $M_1(T,X)$ has been confirmed, that is the conventional BCS theory based on electron-phonon interactions. The mechanism $M_2(T,X)$ is being explored, and several candidates, such as the short-range anti-ferromagnetism spin-correlation mechanism etc. have been proposed. The mechanism, $M_3(T,X)$, is entirely unknown. Because of this we focus on a new superconducting mechanism based on $M_1(T,X)$ and $M_2(T,X)$. The similar idea has also appeared in the literature [5].

In addition, our theory differs from the so-called 'strong coupling' theory which does not have a paring mechanism. Our mechanism is based on the Cooper pairs that are formed by two electrons or two holes through a certain type of interactions caused by mechanical, magnetic or some other more complex interactions. To understand the origin of a Cooper pair, the most important thing is to determine the interaction mediated by a certain media from the background.

## 3. Exploration and Research on Types of Interactions to Form Cooper pairs

In the high-temperature region, we preliminarily consider two single me-chanisms, i.e., $M_2(T,X)$ and $M_1(T,X)$ with $M_2(T,X)$ being the dominant one. Many experiments suggest that the BCS theory cannot fully explain the high-temperature superconductivity. Researchers observed the inhibition for copper-oxides to absorb high frequency photons, which cannot be explained by the BCS theory [6]. Also the BCS theory cannot explain the $T_c$ of cuprates as high as 138K, for instance [7]. However, the BCS theory may still play a role. Lanzara *et al.* [8] have used angle-resolved photoemission spectroscopy to probe electron dynamics, velocity and scattering rates, for three different families of copper oxide superconductors. They observed in all of these materials an abrupt change of electron velocity at 50-80 MeV, which they attributed to electron-phonon coupling involving the movement of the oxygen atoms. This suggests that electron-phonon coupling strongly influences the electron dynamics in the high-temperature superconductors, and must be included in any microscopic theory of superconductivity. There are also some other evidences [9-11]. Therefore, it may be reasonable to assume that the mechanism based on electron-phonon coupling plays merely a secondary role in the high-temperature region.

As has already been pointed out by Zhang [12], the intrinsic actors responsible for high-temperature superconductivity are the concentration of charge carriers and the short-range antiferromagnetism spin correlations among Cu atoms in the $CuO_2$ planes. Although the long-range antiferromagnetism order disappears in the metallic and superconducting phases, strong short-range dynamical spin correlations are observed even at temperatures above 100K [13]. In Fe-based superconductors, the systems also change from paramagnetism to a long-range antiferromagnetisem, doping holes or electrons suppresses both transitions and leads to the appearance of superconductivity similarly [14]. There are enough evidences [15-20] which suggest that the short-range anti-ferromagnetism spin correlations mechanism is the dominant mechanism for the high-temperature superconductivity.

The researches on the weight functions for $M_1(T,X)$ and $M_2(T,X)$ are being carried out.

## 4. Summary and Conclusion

To solve the problem in the introduction, we have proposed a unified mixed-mechanism for the superconductivity [4]. This mixed-mechanism depends on several single mechanisms other than a single mechanism. In our mixed-mechanism, the dominant mechanism varies with the change of temperature region and type of materials. By using the mixed-mechanism, many problems which can not be understood by using a single mechanism can be explained. The further studies are being conducted.

## Acknowledgements

This work was supported by the Foundation of State Key Laboratory of Mechanical Transmission, Chongqing University (SKLT-KFKT-200908).

The authors would like to thank Dr. Shui-Quan Deng and Dr. Xing-Gang Wu for helpful discussions

The authors would like to thank Dr.M.de Llano for helpful direction exceptionally.

Note appended: This paper exchanged on the 2011 New3SC-8, chongqing china , and passed the Review.